\documentstyle[pra,aps,multicol,eqsecnum]{revtex}
\def\be{\begin{eqnarray}}
\def\ee{\end{eqnarray}}
\def\l{\langle}
\def\r{\rangle}   

\draft
\begin{document}

\title{Quantum information distributors: Quantum network for symmetric and 
asymmetric cloning in arbitrary dimension and continuous limit}

\author{
Samuel L.\ Braunstein${}^{1}$, Vladim\'{\i}r Bu\v{z}ek${}^{2}$\cite{padd}
and Mark Hillery${}^{3}$}
\address{
${}^1$Informatics, University of Wales, Bangor LL57 1UT, UK \\
${}^{2}$Department of Physics, University of Queensland, QLD 4072,
Brisbane, Australia\\
${}^{3}$Department of Physics and Astronomy, Hunter College of CUNY, 695,
Park
Avenue, New York, NY 10021, U.S.A.} 

\date{5 September 2000}

\maketitle

\begin{abstract}
We show that for any Hilbert-space dimension, the optimal universal 
quantum cloner can be constructed from essentially the same quantum 
circuit, i.e., we find a universal design for universal cloners.
In the case of infinite dimensions (which includes continuous variable 
quantum systems) the universal cloner reduces to an essentially 
classical device. More generally, we construct a universal quantum 
circuit for distributing qudits in any dimension which acts
covariantly under generalized displacements and momentum kicks.
The behavior of this covariant distributor is controlled by its initial
state. We show that suitable choices for this initial state yield
both universal cloners and optimized cloners for limited alphabets of
states whose states are related by generalized phase-space displacements.
\end{abstract}

\pacs{03.67.-a, 03.65.Bz}

\begin{multicols}{2}

\section{Introduction}
One of the main tasks in quantum information processing and quantum
computing is the {\em distribution} of quantum information encoded in 
the states of quantum systems. Assume a quantum system labelled as $1$ 
is prepared in an {\em unknown} pure state described by a state vector
$|\Psi\r_1$ in an $N$-dimensional Hilbert space.  The task is to transfer 
{\em partially} the information encoded in system $1$ into a second system 
in a covariant way. That is, the fidelity of the operation should not 
depend on the particular choice of the input state $|\Psi\r_1$.
In addition, we want to control the amount of information transferred from 
system $1$ to system $2$. 
One of the simplest examples of such a transformation
is state-swapping, when the state of system $1$ is swapped
with the (known) state of system $2$. In this case the complete
information is transferred. Another option is to leave the system
$1$ in the original state. These two operations can be performed with 
unit fidelity irrespective of the input state of system $1$. We can also 
consider a case intermediate between these two limiting cases, 
i.e., between no transfer and the complete transfer of information. 
One interesting version
of this intermediate transformation involves the copying (cloning) of 
quantum information from system $1$ to system $2$, where,
after the transformation,  
each of the systems $1$ and $2$ has the same reduced state, which is
itself as close as possible to the original state $|\Psi\r_1\langle \Psi|$.
In this case we often require that the fidelity of the 
information transfer does not depend on the initial state. 

It is now well 
known that quantum information cannot be exactly copied \cite{wootters}. 
This {\em no-cloning} theorem has important consequences for the whole 
field of quantum information processing \cite{nielsen}. Nonetheless, 
considerable progress has been made in the development of approximate 
quantum cloners.  Most of the effort has focused on 
two types of these, universal cloners that 
copy all input states equally well \cite{buzek1}, and probabilistic cloners 
that copy a known set of states perfectly, but do so with a probability 
which is less than one \cite{duan}. 
In what follows we shall concentrate on {\em universal} devices.

Quantum cloning both illuminates the limits imposed by
quantum mechanics on the manipulation of quantum information
and can be useful in applications.  It has been shown to be 
useful in improving the performance of imperfect
quantum detectors \cite{deuar} and in improving the performance
of certain quantum computations \cite{galvao}. 
In addition, it has been shown in Ref.~\cite{Gisin99b} that  quantum cloners
can be used as optimal eavesdropping devices on the 6-state cryptographic
protocol.
We should also note that recently
an interesting cloning  experiment has been proposed \cite{simon}.
Moreover, two experiments have been independently reported
\cite{demartini,li} this year  in which cloning of optical fields
has been realized.

Universal cloners can be either symmetric or asymmetric.  In a
symmetric cloner the quantum information is divided equally and
the output clones are identical.  In an asymmetric cloner one of the 
clones receives more of the input quantum information than the
other.  Symmetric cloners were first developed to copy qubits \cite{buzek1,gisin},
but have been extended to copy states in spaces of arbitrary dimension 
\cite{buzek2,Cerf2000,Albeverio2000}, and it has been proven that these
cloners are optimal \cite{bruss,gisin,werner}.  The study of 
asymmetric cloners also began with the consideration of qubits
\cite{Cerf98,buzek3,niu} and has been recently extended to
systems of arbitrary dimension \cite{Cerf2000}.  What we shall
do here is to exhibit a quantum circuit for symmetric
and asymmetric cloners in arbitrary numbers of dimensions.
In order to emphasize that what these devices do is distribute
quantum information, we shall refer to them as quantum 
information distributors (QID's).  The circuit consists of
four controlled-NOT gates, or rather their generalization to
$N$ dimensions, and its form is the same for any number of
dimensions.  There are two inputs to this circuit. The first is
the state which supplies the information to be distributed 
between the two outputs. The second acts as a program and 
determines how the information is distributed.  The     	 infinite-dimensional version of this circuit 
allows us to describe quantum information distributors 
for continuous variables.

Let us formulate our problem more exactly. Assume the original
quantum system is in a pure state 
\be
|\Psi\r_1 = \sum_{n=0}^{N-1}c_n|x_n\r \;.
\label{1.1}
\ee
At the output of
the quantum information distributor we would like to have two quantum
systems each with a state described in a {\em covariant} form
\be
\hat{\rho}_1^{(\rm out)}&=&(1-\beta^2)|\Psi\r_1\l\Psi | + \frac{\beta^2}{N}
\hat{\openone}_1 
\nonumber
\\
\hat{\rho}_2^{(\rm out)}&=&(1-\alpha^2)|\Psi\r_2\l\Psi | + \frac{\alpha^2}{N}
\hat{\openone}_2 \;,
\label{1.2}
\ee
where the real parameters $\alpha$ and $\beta$ quantify the amount of
information which has been transferred from one system to the other. 
In particular, if $\beta=0$, then {\em no} information has been transfered
from the original system, while if $\alpha=0$, then all of the information
in system $1$ has been transferred to system $2$.
The parameters $\alpha$ and $\beta$ are related (see below).
From the covariant form of the output density operators it follows
that the fidelity of the 
information transfer is input-state independent.
The terms proportional to $\hat{\openone}/N$ in the density operators 
describe the amount of noise introduced into the systems at the
output by the information transfer process. 

Our task in this paper is to develop a quantum circuit for the 
universal quantum information distributor for arbitrary-dimensional 
quantum systems. 
In Sec.~\ref{sec2} we start our discussion with the mathematical formalism 
needed to investigate our problem. Then in Section~\ref{sec3}
we present a quantum network for the universal quantum information 
distributor, while in Section~\ref{sec4} we generalize the discussion
to continuous variables. Finally, in Section~\ref{sec5} we summarize
our results.

\section{From discrete to continuous variables}
\label{sec2}
In order to make the discussion self-contained we first present a 
brief review of the formalism describing quantum states in a 
finite-dimensional Hilbert space. Here we follow the notation introduced 
in Refs.~\onlinecite{Buz92,Opat95} (see also Ref.~\onlinecite{Enk99}).
Let the $N$-dimensional Hilbert space be spanned by $N$ orthogonal
normalized vectors $|x_{k}\rangle$ and equivalently by $N$ vectors
$|p_{l}\rangle$, $k, l = 0, \ldots, N-1$, where these bases are related 
by the discrete Fourier transform
\begin{eqnarray}
\label{2.1}
|x_{k}\rangle &=& \frac{1}{\sqrt{N}} \sum_{l=0}^{N-1} \exp \Bigl(
-i\frac{2\pi}{N}kl \Bigr) |p_{l}\rangle 
\nonumber \\
|p_{l}\rangle &=& \frac{1}{\sqrt{N}} \sum_{k=0}^{N-1} \exp \Bigl(
i\frac{2\pi}{N}kl \Bigr) |x_{k}\rangle \;.
\end{eqnarray}
Without loss of generality, it can be assumed that these bases are sets 
of eigenvectors of non-commuting operators $\hat X$ and $\hat P$:
\begin{eqnarray}
\label{2.2}
\hat X |x_{k}\rangle = k|x_{k}\rangle \;, \quad
\hat P |p_{l}\rangle = l|p_{l}\rangle \;,
\end{eqnarray}
that is,
\begin{eqnarray}
\label{2.3}
\hat X &=& \sum_{k=0}^{N-1} k |x_{k}\rangle \langle x_{k}| 
\nonumber\\
\hat P &=& \sum_{l=0}^{N-1} l |p_{l}\rangle \langle p_{l}| \;.
\end{eqnarray}
For instance, we  can assume that the operators $\hat X$ and $\hat P$
are related to a
discrete position and momentum of a particle on a ring with a finite 
number of equidistant sites \cite{PB}. Specifically, we can introduce 
a length scale, $L$, and two operators, the position 
$\hat x$ and the momentum $\hat p$, such that
\begin{eqnarray}
\label{2.4}
\hat x |x_{k}\rangle = x_k|x_{k}\rangle \;, \quad
\hat p |p_{l}\rangle = p_l|p_{l}\rangle \;,
\end{eqnarray}
where
\begin{eqnarray}
\label{2.5}
x_k= L 
\sqrt{\frac{2\pi}{N}} k ;
\qquad
p_l=\frac{\hbar}{L} \sqrt{\frac{2\pi}{N}} l\;.
\end{eqnarray}
The length, $L$ can, for example, be taken equal to $\sqrt{\hbar/\omega m}$, 
where $m$ is the mass and $\omega$ the frequency of 
a quantum ``harmonic''oscillator within a finite dimensional 
Fock space (in what follows we use units such that
$\hbar=1$).

The squared absolute
values of the scalar product of eigenkets (\ref{2.2}) do not depend on the
indices $k$, $l$:
\begin{eqnarray}
\label{2.6}
|\langle x_{k}|p_{l} \rangle |^{2} = 1/N \;,
\end{eqnarray}
which  means that pairs $(k,l)$ form a discrete phase space 
(i.e., pairs $(k,l)$ represent ``points'' of the discrete phase space) 
on which a Wigner function can be defined \cite{Woot87}.
Next we introduce operators which shift (cyclicly permute)
the basis vectors \cite{GTP88}:
\begin{eqnarray}
\label{2.7}
\hat R_{x}(n)|x_{k}\rangle &=& |x_{(k+n){\rm mod}\, N}\rangle \nonumber \\
\hat R_{p}(m)|p_{l}\rangle &=& |p_{(l+m){\rm mod}\, N}\rangle\;,
\end{eqnarray}
where the sums of indices are taken modulo $N$ (this summation rule
is considered  throughout this paper, where it is clear we will not
explicitly write the symbol ${\rm mod}\, N$).
The operators  $\hat R_{x}(n)$ and	$\hat R_{p}(m)$ can be 
expressed as powers of  the operators $\hat R_{x}(1)$ and 
$\hat R_{p}(1)$, respectively:
\begin{eqnarray}
\label{2.8}
\hat R_{x}(n) = \hat R_{x}^{n}(1)\;, \quad
\hat R_{p}(m) = \hat R_{p}^{m}(1)\;.
\end{eqnarray}
In the $x$-basis these operators can be expressed as
\begin{eqnarray}
\label{2.9}
\langle x_{k}|\hat R_{x}(n)|x_{l} \rangle &=& \delta _{k+n,l}
\nonumber \\
\langle x_{k}|\hat R_{p}(m)|x_{l} \rangle &=& \delta _{k,l} \exp \Bigl(
i \frac{2 \pi}{N} ml \Bigr) \;.
\end{eqnarray}
Moreover these operators fulfill the Weyl commutation relation 
\cite{Weyl,Sant,Stov}
\begin{eqnarray}
\label{2.10}
\hat R_{x}(n) \hat R_{p}(m) = \exp \Bigl( i \frac{2\pi}{N} mn  \Bigr)
\hat R_{p}(m) \hat R_{x}(n) \;;
\end{eqnarray}
although they do not commute, they form a representation of an Abelian
group in a ray space. We can displace a state in arbitrary order using
$\hat R_{x}(n) \hat R_{p}(m)$ or $\hat R_{p}(m) \hat R_{x}(n)$, the
resulting state will be the same --- the corresponding kets will differ
only by an unimportant multiplicative factor. We see that  
the operators $\hat R_{x}(n)$ and $\hat R_{p}(m)$ 
displace states in the directions $x$ and $p$, respectively.
The product
$\hat R_{x}(n) \hat R_{p}(m)$ acts as a displacement operator in the
discrete phase space $(k, l)$ \cite{Buz95}. These operators can be 
expressed via the generators of translations (shifts)
\begin{eqnarray}
\label{2.11}
\hat R_{x}(n) &=& \exp \Bigl( -i \frac{2\pi}{N} n \hat P  \Bigr) =
\exp(-i x_n \hat{p})
\nonumber
\\
\hat R_{p}(m) &=& \exp \Bigl( i \frac{2\pi}{N} m \hat X  \Bigr) =
\exp(i p_m \hat{x})\;. 
\end{eqnarray}       
We note that the structure of the group associated with the
operators $\hat R_{x}(n)$ and $\hat R_{p}(m)$ is reminiscent of the
group of phase-space translations (i.e., the Heisenberg group)
in quantum mechanics \cite{Fivel95}.

A general single-particle state in the $x$-basis can be expressed as
\begin{eqnarray}
\label{2.12}
|\Psi\rangle_1 =\sum_{k=0}^{N-1} c_k
|x_k\rangle_1\;; \qquad \sum_{k=0}^{N-1} |c_k|^2 =1\;.
\end{eqnarray}       
The basis of maximally entangled two-particle states (the analogue
of the Bell basis for spin-$\case{1}{2}$ particles) can be written as
\cite{Fivel95}
\begin{eqnarray}
\label{2.13}
|\Xi_{mn}\rangle = \frac{1}{\sqrt{N}} \sum_{k=0}^{N-1} \exp \Bigl(
i\frac{2\pi}{N} mk \Bigr) |x_{k}\rangle|x_{(k-n){\rm mod}\,N}\rangle \,,\!\!\!
\end{eqnarray}
where $m,n=0,\dots,N-1$. We can also rewrite these maximally entangled
states in the $p$-basis:
\begin{eqnarray}
\label{2.14}
|\Xi_{mn}\rangle = \frac{1}{\sqrt{N}} \sum_{l=0}^{N-1} \exp \Bigl(
-i\frac{2\pi}{N} nl \Bigr) |p_{(m-l){\rm mod}\,N}\rangle|p_{l}\rangle\;.\!\!\!
\end{eqnarray}
The states $|\Xi_{mn}\rangle$ form an orthonormal basis
\begin{eqnarray}
\label{2.15}
\langle \Xi_{kl}|\Xi_{mn}\rangle = \delta_{k,m}\delta_{l,n} \;,
\end{eqnarray}
with
\begin{eqnarray}
\label{2.16}
\sum_{m,n=0}^{N-1} |\Xi_{mn}\rangle\langle\Xi_{mn}| = \hat{\openone}\otimes
\hat{\openone}\;.
\end{eqnarray}
In order to prove the above relations we have used the standard relation
$\sum_{n=0}^{N-1}\exp[2\pi i (k-k') n/N] = N \delta_{k,k'}$.

It is interesting to note that the whole set of $N^2$ maximally entangled
states $|\Xi_{mn}\rangle$ can be generated from the state
$|\Xi_{00}\rangle_{23}$ by the action of {\em local} unitary operations
(shifts) of the form 
\begin{eqnarray}
\label{2.17}
|\Xi_{mn}\rangle_{23} = 
\hat{\openone}_2\otimes \hat{R}_x(n) \hat{R}_p(m) |\Xi_{00}\rangle_{23} \;,
\end{eqnarray}
acting just on system $3$ in this particular case.

From the definition of the states $|\Xi_{mn}\rangle_{23}$ it follows that 
they are simultaneously eigenstates of the operators $\hat{X}_2 - \hat{X}_3$
and $\hat{P}_2 + \hat{P}_3$:
\begin{eqnarray}
\label{2.18}
(\hat{X}_2 - \hat{X}_3)|\Xi_{mn}\rangle_{23} &=& n |\Xi_{mn}\rangle_{23}
\nonumber
\\
(\hat{P}_2 + \hat{P}_3)|\Xi_{mn}\rangle_{23} &=& m |\Xi_{mn}\rangle_{23} \;.
\end{eqnarray}       
We easily see that for $N=2$ the above formalism reduces to the well-known
spin-$\case{1}{2}$ particle (qubit) case.

Now we introduce generalizations of the two-qubit C-NOT gate 
(see also Ref.~\onlinecite{alber}).  In the case of qubits the C-NOT gate 
is represented by a two-particle operator such that if the first 
(control) particle labelled $a$ is in the state $|0\rangle$ nothing 
``happens'' to the state of the second (target) particle labelled $b$.
If, however, the control particle is in the state $|1\rangle$ 
then the state of the target is ``flipped'', i.e., the state $|0\rangle$ 
is changed into the state $|1\rangle$ and vice versa. Formally we can 
express the action of this C-NOT gate  as a two-qubit operator of the form
\begin{eqnarray}
\label{2.19}
\hat{D}_{ab}=\sum_{k,m=0}^1 |k\rangle_a\langle k|\otimes
|(m+k){\rm mod}\,2\rangle_b\langle m|\;. 
\end{eqnarray}       
We note that in principle one can introduce an operator 
$\hat{D}^\dagger_{ab}$ defined as
\begin{eqnarray}
\label{2.20}
\hat{D}^\dagger_{ab}=\sum_{k,m=0}^1 |k\rangle_a\langle k|\otimes
|(m-k){\rm mod}\,2\rangle_b\langle m|\;. 
\end{eqnarray}       
In the case of qubits these two operators are equal. This is not the case
when the dimension of the Hilbert space is larger than 2 \cite{alber}. Let 
us generalize the above definition of the operator $\hat{D}$ for $N>2$. 
Before doing so, we shall simplify our notation. Because we will work 
mostly in the $x$-basis we shall use the notation 
$|x_k\rangle\equiv|k\rangle$ where it may be done so unambiguously. With 
this in mind we now write
\begin{eqnarray}
\label{2.21}
\hat{D}_{ab}=
\sum_{k,m=0}^{N-1} |k\rangle_a\langle k|\otimes
|(m+k){\rm mod}\,N\rangle_b\langle m|\;. 
\end{eqnarray}       
From the definition (\ref{2.21}) it follows that the operator
$\hat{D}_{ab}$ acts on the basis vectors as
\be
\hat{D}_{ab}|k\r|m\r = |k\r|(k+m) {\rm mod}\, N\r\;,
\label{2.22}
\ee
which means that this operator is equal to the conditional adder
\cite{vedral96,pittenger} and can be performed with the help of
a simple quantum network as discussed in \cite{vedral96}.

If we take into account the definition of the shift operator
$\hat{R}_x(n)$ given by Eq.~(\ref{2.7}) 
and the definition of the position and momentum operators $\hat{x}$
and $\hat{p}$ given by Eq.~(\ref{2.8}) 
we can rewrite the operator
$\hat D_{ab} $ as:
\begin{eqnarray}
\label{2.23}
\hat{D}_{ab}&=&\sum_{k,m=0}^{N-1} |k\rangle_a\langle k|\otimes 
\hat{R}_x^{(b)}(k)|m\rangle_b\langle m| 
\\
\nonumber
&\equiv& 
\sum_{k=0}^{N-1} |k\rangle_a\langle k|\otimes 
\hat{R}_x^{(b)}(k) = e^{-i\hat{x}_a\hat{p}_b} \;, 
\end{eqnarray}       
and analogously
\begin{eqnarray} 
\label{2.24}
\hat{D}_{ab}^\dagger&=&
\sum_{k,m=0}^{N-1} |k\rangle_a\langle k|\otimes
|(m-k){\rm mod}\,N\rangle_b\langle m| 
\\
\nonumber
&\equiv& 
\sum_{k=0}^{N-1} |k\rangle_a\langle k|\otimes 
\hat{R}_x^{(b)}(-k) = e^{i\hat{x}_a\hat{p}_b}\;, 
\end{eqnarray}     
where the superscripts $a$ and $b$ indicate on which Hilbert space the
given operator acts.
Now we see that for $N>2$ 
the two operators $\hat D$ and $\hat{D}^\dagger$ do differ;
they describe conditional shifts in opposite directions.
We see that the generalization of the
C-NOT operator are the {\em conditional shifts}. 
The amount by which the target
(in our case particle $b$) is shifted 
depends on the state of the control particle ($a$). 

\subsection{Continuous limit}
In the $N\rightarrow\infty$ limit we have to take special
care in handling the expressions for the eigenstates of the position
and momentum  operators \cite{Dirac}. To avoid divergences
we have to regularize our states by ``smearing'' them. In other
words, the eigenstate of the operator $\hat{x}$ is replaced
by a squeezed displaced 
state (see e.g., Ref.~\onlinecite{Buz95}) with reduced quadrature
fluctuations in the $\hat{x}$ direction (see below). To express these
states explicitly we utilize the Wigner function representation, which
for pure states is defined as
\be
W_{_{|\Psi\r}}(x,p) = \frac{1}{\sqrt{2\pi}} \int_{-\infty}^{\infty}
\!dz \, \varphi_{_{|\Psi\r}}(x_-) \, \,
\varphi_{_{|\Psi\r}}^*(x_+) 
\, e^{ipz} \;,\!\!\!
\label{2.25}
\ee
where $x_\pm=(x\pm{z}/{2})$ and 
$\varphi_{_{|\Psi\r}}(x)$ is the wave function of the state $|\Psi\r$,
i.e., $\varphi_{_{|\Psi\r}}(x)=\l x|\Psi\r$, which in the $|x\r$ basis is
expressed as
\be
|\Psi\r=\frac{1}{\sqrt{2\pi}}\int_{-\infty}^{\infty}
dx\, 
\varphi_{_{|\Psi\r}}(x) |x\r\;,
\label{2.26}
\ee
where we have used the relation $\l x| y\r = \sqrt{2\pi}\, \delta(x-y)$.
The Wigner function $W_{|\Psi\r}(x,p)$ is a quasi-probability 
distribution in phase space and is normalized so that
\be
\frac{1}{2\pi}
\int_{-\infty}^{\infty} dx\, dp \,  W_{_{|\Psi\r }}(x,p) = 1\;,
\label{2.27}
\ee
where $dx\,dp/ 2\pi$ is the invariant measure in phase space (here we have 
taken $\hbar=1$).

With these definitions we can represent a regularized version of
an  eigenstate of the position operator $\hat{x}$ 
with mean value equal to zero, $|x_0\r$,  
as a state described by the Gaussian Wigner function:
\be
W_{_{|x_0\r}}(x,p) = 2 \exp( -e^{2\xi} x^2 - e^{-2\xi} p^2)\;,
\label{2.28}
\ee
for which the variances of the position and momentum operators are
$(\Delta\hat{x})^2 =\frac{1}{2} e^{-2\xi}$ and
$(\Delta\hat{p})^2 =\frac{1}{2} e^{2\xi}$, respectively. 
The state (\ref{2.28}) is a minimum uncertainty state, i.e.,
$(\Delta\hat{x})(\Delta\hat{p})=\case{1}{2}$ irrespective of the value
of the squeezing parameter $\xi$.
For the mean
excitation number we find the expression $\bar{n}=\sinh^2 \xi$. We see
that in the limit $\xi\rightarrow\infty$,
the state described by Wigner function (\ref{2.28})
is indeed a state with no fluctuations in the $x$ direction
at the expense of infinite fluctuations in the $p$ direction.
In other words in the limit $\xi\rightarrow\infty$ the state
(\ref{2.28}) is an eigenstate of $\hat{x}$.

Analogously  a regularized  
 eigenstate of the momentum operator is described by the Wigner function
\be
W_{_{|p_0\r}}(x,p) = 2 \exp( -e^{-2\xi} x^2 - e^{2\xi} p^2)\;,
\label{2.29}
\ee
where for the variances of the position and momentum operators we find
$(\Delta\hat{x})^2 =\frac{1}{2} e^{2\xi}$ and
$(\Delta\hat{p})^2 =\frac{1}{2} e^{-2\xi}$, respectively. 

The wave functions corresponding to the states (\ref{2.28}) and (\ref{2.29})
read
\be
\varphi_{_{|x_0\r}}(x) 
= 2^{1/4} e^{\xi/2} \exp\Bigl(-\frac{e^{2\xi} x^2}{2}
\Bigr)\;,
\label{2.30}
\ee
and
\be
\varphi_{_{|p_0\r}}(x) = 2^{1/4} e^{-\xi/2} 
\exp\Bigl(-\frac{e^{-2\xi} x^2}{2}
\Bigr)\;,
\label{2.31}
\ee
respectively. We denote the corresponding ket vectors
as $|x_0(\xi)\r$ and $|p_0(\xi)\r$, where we have explicitly 
indicated that these states are regularized versions of 
two specific eigenstates of the position and momentum operators.
The $x$-distribution of the state $\varphi_{_{|x_0\r}}(x)$
is defined as usual, i.e., $P_{_{|x_0\r}}(x)
=|\varphi_{_{|x_0\r}}(x)|^2$ and is normalized to unity as
$\int dx\, P_{_{|x_0\r}}(x)/\sqrt{2\pi}=1$. This distribution can
also be obtained from the Wigner function (\ref{2.28}) via 
integration over $p$, i.e.,
\be
P_{_{|x_0\r}}(x) &=& \frac{1}{\sqrt{2\pi}} \int dp\, 
W_{_{|x_0\r}}(x,p)
\nonumber
\\ 
&=& 2^{1/2} e^{\xi} \exp(-e^{2\xi} x^2 )\;,
\label{2.32}
\ee
which in the large $\xi$ limit gives 
$\lim_{\xi\rightarrow\infty} P_{_{|x_0\r}}(x) = \sqrt{2\pi}\, \delta(x)$,
as expected.

In an analogous way we define a maximally entangled two-mode
state $|\Xi_{00}(\xi)\r$ in the continuous limit. Specifically, we define
this state in a regularized form for which the Wigner function
reads \cite{Braunstein98}

\end{multicols}
\vspace{-0.2cm}
\noindent\rule{0.5\textwidth}{0.4pt}\rule{0.4pt}{0.6\baselineskip}
\vspace{0.2cm}
\be
W_{_{|\Xi_{00}\r}}(x_1,p_1;x_2,p_2) = 
4 \exp\left\{ 
-\frac{e^{2\xi}}{2}\left[(x_1-x_2)^2 + (p_1+p_2)^2\right]
-\frac{e^{-2\xi}}{2}\left[(x_1+x_2)^2 + (p_1-p_2)^2\right]
\right\} \;.
\label{2.33}
\ee
  \hfill\noindent\rule[-0.6\baselineskip]%
  {0.4pt}{0.6\baselineskip}\rule{0.5\textwidth}{0.4pt}
\vspace{-0.2cm}
\begin{multicols}{2}
\noindent
This is a Wigner function describing a two-mode squeezed vacuum. If 
we trace over one of the modes, i.e., if we perform an integration over
the parameters $x_2$ and $p_2$ we obtain from (\ref{2.33}) a Wigner 
function of a thermal field
\be
W_{_{\rm th}}(x_1,p_1) &=&\frac{1}{2\pi}\int_{-\infty}^{\infty} 
dx_2\, dp_2
W_{_{|\Xi_{00}\r}}(x_1,p_1;x_2,p_2) 
\nonumber
\\
&=& 
\frac{2}{1+2\bar{n}}
\exp\Bigl(-\frac{x_1^2 + p_1^2}{1+2\bar{n}}\Bigr)\;,
\label{2.34}
\ee
where $\bar{n}=\sinh^2\xi$ is the mean excitation number in the two-mode
squeezed vacuum under consideration. We note that the thermal state
(\ref{2.34}) is a maximally mixed state (i.e., with the state with the
highest value of the von Neumann entropy) for a given mean excitation
number. This means that the pure state (\ref{2.33}) is the most 
entangled state for a given mean excitation number.
From this it follows that to create a truly maximally entangled state,
i.e., the state (\ref{2.33}) in the limit $\xi\rightarrow\infty$, an 
infinite number of quanta is needed and so infinite energy.

The two-mode wave function of the state (\ref{2.33}) in the 
$x$-representation reads
\be
\varphi_{_{|\Xi_{00}\r}}(x_1;x_2) 
= \sqrt{2}  \exp\Bigl(
-\frac{e^{2\xi}}{4}\tilde{x}_-^2
-\frac{e^{-2\xi}}{4}\tilde{x}_+^2
\Bigr)\;, 
\label{2.35}
\ee
where $\tilde{x}_{\pm}=(x_1\pm x_2)$.
In what follows we shall denote this regularized version of the
maximally entangled state in a semi-infinite Hilbert space 
as $|\Xi_{00}(\xi)\rangle$. 
Now that we have laid out the formalism, we can resume our 
discussion.

\section{Network for quantum information distribution}
\label{sec3}
We have shown earlier \cite{buzek1,buzek2,Buzek97} that to perform 
quantum cloning we need apart from systems $1$ and $2$,
between which the information is shared, an additional 
quantum system $3$ which mediates 
the distribution of the quantum information. Following this philosophy, 
we assume a quantum information distributor to be a two-particle system 
($2$ and $3$) each of the same physical type as the original system $1$. 
Let us assume that the quantum distributor is initially prepared in the 
most general two-particle pure state 
\begin{eqnarray} 
\label{3.1}
|\Phi\rangle_{23} = 
\sum_{m,k=0}^{N-1} d_{mk} |m\rangle_2 |k\rangle_3\;.
\end{eqnarray}     

In analogy with the quantum computational network used in the quantum 
cloner \cite{Buzek97} we assume the QID network to be
\be
\hat{U}_{123}=
\hat{D}_{31}\hat{D}_{21}^\dagger\hat{D}_{13}\hat{D}_{12}\;,
\label{3.2}
\ee
with the idea being that the flow of information in the quantum distributor,
as described by the unitary operator (\ref{3.2}), is governed by the 
preparation of the distributor itself, i.e., by the choice of the 
state (\ref{3.1}). In other words, we imagine the transformation 
(\ref{3.2}) as a universal  ``processor'' or distributor and the 
state (\ref{3.1}) as ``software'' 
through which the information flow is controlled.
Using relation (\ref{2.23}) we can rewrite the QID transformation
as (see also Ref.~\onlinecite{Cerf1999})
\be
\hat{U}_{123}=\exp[-i (\hat{x}_3-\hat{x}_2)\hat{p}_1]
 \exp[-i \hat{x}_1(\hat{p}_2+\hat{p}_3)]\;.
\label{3.3}
\ee

The distribution
of information encoded in the original particle is performed 
via a sequence
of four conditional shifts $\hat{D}$. The output state 
of the three particle system after the four controlled 
shifts are applied is 
\begin{eqnarray} 
\label{3.4}
|\Omega\rangle_{123}= 
\hat{D}_{31}\hat{D}_{21}^\dagger\hat{D}_{13}\hat{D}_{12}
|\Psi\rangle_1|\Phi\rangle_{23}\;.
\end{eqnarray}     
The four operators $\hat{D}$ act on the basis vectors 
$|n\rangle_{1}|m\rangle_{2}|k\rangle_{3}$ as
\end{multicols}
\vspace{-0.2cm}
\noindent\rule{0.5\textwidth}{0.4pt}\rule{0.4pt}{0.6\baselineskip}
\vspace{0.2cm}
\begin{eqnarray} 
\label{3.5}
\hat{D}_{31}\hat{D}_{21}^\dagger\hat{D}_{13}\hat{D}_{12}
|n\rangle_{1}|m\rangle_{2}|k\rangle_{3}=
|(n-m+k){\rm mod}\,N\rangle_{1}\,|(m+n){\rm mod}\,N\rangle_{2}
\,|(k+n){\rm mod}\,N\rangle_{3}\;.
\end{eqnarray}     
  \hfill\noindent\rule[-0.6\baselineskip]%
  {0.4pt}{0.6\baselineskip}\rule{0.5\textwidth}{0.4pt}
\vspace{-0.2cm}
\begin{multicols}{2}
\noindent
As we shall see, the choice of the state $|\Phi\rangle_{23}$
controls the flow of the quantum information contained in the state
$|\Psi\rangle_{1}$ through the QID.  

Before examining this issue, however, it is useful to explore the 
covariance properties of this distributor for any choice of 
$|\Phi\rangle_{23}$.  A device is covariant with respect to the 
transformation $\hat{U}$, if application of $\hat{U}$ to the input, 
i.e., $|\Psi\rangle\rightarrow
\hat{U}|\Psi\rangle$ implies that the output density matrix representing
the pair of outputs, $\hat{\rho}^{(\rm out)}$, transforms as \cite{gisin2}
\begin{equation}
\rho^{(\rm out)} \rightarrow \hat{U}\otimes \hat{U} 
\rho^{(\rm out)} \hat{U}^{-1}\otimes \hat{U}^{-1} \;.
\end{equation}
When examining whether the distributor is covariant with respect to 
transformations of the form $\hat R_x(n)\hat R_p(n)$,
it is sufficient to confirm this covariant action for `displacements'
along the $x$- and $p$-axis separately, given by $\hat R_x(n)$ and
$\hat R_p(n)$, respectively. If the state to be distributed is 
$\hat R_x(n)|\Psi\rangle_{1}$ we find
\begin{equation}
\hat R_x(n)_1|\Psi\rangle_{1}|\Phi\rangle_{23}
\rightarrow 
\hat R_x(n)_1 \hat R_x(n)_2 \hat R_x(n)_3
|\Omega\rangle_{123} \;,
\end{equation}
where $|\Omega\rangle_{123}$ is given by Eq.~(\ref{3.4}).  Similarly, if 
the input state is $\hat R_p(n)|\Psi\rangle_{1}$, we have
\begin{equation}
\hat R_p(n)_1|\Psi\rangle_{1}|\Phi\rangle_{23}
\rightarrow 
\hat R_p(n)_1 \hat R_p(n)_2 \hat R_p(-n)_3
|\Omega\rangle_{123} \;.
\end{equation}
Combinations of these two `displacements' act in the natural way, so
that if we `translate' the input state by a certain amount, the
reduced density matrixes of the three outputs are translated by the
same amount, and if we perform a momentum `translation' on the input 
state, the reduced density matrixes of outputs $1$ and $2$ are 
translated in momentum by the same amount, while that of output $3$ has
its momentum translated by the opposite amount.  This implies that 
this QID is covariant with respect to translations and momentum 
translations, and that the fidelities of the output reduced density 
matrixes are unaffected when these transformations are applied to 
the input.

Having established the covariant action of our distributor in arbitrary
dimensions for any input state $|\Phi\rangle_{23}$,
we now wish to determine how this state affects the flow of
quantum information in the QID.
\hfill\break
{\bf (i)} Let us first assume that the QID state $|\Phi\r$ is initially
prepared in the maximally entangled state $|\Xi_{00}\r_{23}$ given
by Eq.~(\ref{2.14}). Taking the original system to be prepared in the state
(\ref{1.1}) we find after the QID transformation 
\be
\hat{U}_{123}|\Psi\r_1|\Xi_{00}\r_{23} = 
|\Psi\r_1|\Xi_{00}\r_{23} \;,
\label{3.6}
\ee
that system $1$ remains in the original state while the QID remains in 
its initial maximally entangled state $|\Xi_{00}\r_{23}$. This means 
that even though the three-particle system has interacted via four 
controlled shifts the total state is unchanged. 
\hfill\break
{\bf (ii)} Instead, let us assume the QID is initially prepared in the
product state
\be
|\Phi\r_{23} = |x_0\r_2 |p_0\r_3\;,
\label{3.7}
\ee
where $|p_0\r$ is an eigenstate of the momentum operator with the mean 
value equal to zero. At the output we then find
\be
\hat{U}_{123}|\Psi\r_1|x_0\r_2 |p_0\r_3 = 
|\Psi\r_2|\Xi_{00}\r_{13}\;,
\label{3.8}
\ee
which means that the information from the system $1$ is completely
transfered to the system $2$  while at the output the system 
$1$ and $3$ are  in the maximally entangled state 
$|\Xi_{00}\r_{13}$.  Note that the output here is a state-swapped
version of the output in Eq.\ (\ref{3.6}).

Since these two cases realize the two extreme situations (no 
information transfer and complete information transfer) it is natural 
to ask what is the action of the QID if it is prepared in a linear 
superposition of the states $|\Xi_{00}\r_{23}$ and $|x_0\r_2 |p_0\r_3$. 
Let us take the input state of the QID to be
\be
|\Phi\r_{23}=
\alpha|\Xi_{00}\r_{23}+\beta |x_0\r_2 |p_0\r_3\;,
 \label{3.9}
\ee
where $\alpha$ and $\beta$ are real parameters. Note, that from the 
normalization condition $\l \Phi|\Phi\r=1$ it follows that these 
parameters must fulfill the condition
\be
\alpha^2 +\beta^2 +\frac{2\alpha\beta}{N}=1\;.
\label{3.10}
\ee
When the QID transformation is applied  with the QID initially
prepared in the state (\ref{3.9}) the output state becomes
\be
\hat{U}_{123}|\Psi\r_1|\Phi\r_{23} = 
\alpha|\Psi\r_1|\Xi_{00}\r_{23}+\beta |\Psi\r_2|\Xi_{00}\r_{13}\;.
\label{3.11}
\ee
Tracing over the systems $2$ ($1$) and $3$ we find the reduced stated for
system $1$ ($2$) at the output to be described by
\be
\hat{\rho}_1^{(\rm out)}&=&(\alpha^2+\frac{2\alpha\beta}{N})
\hat{\rho}^{({\rm in})}
+ \frac{\beta^2}{N}
\hat{\openone}\, ;
\nonumber
\\
\hat{\rho}_2^{(\rm out)}&=&(\beta^2+\frac{2\alpha\beta}{N})
\hat{\rho}^{({\rm in})}
+ \frac{\alpha^2}{N}
\hat{\openone} \, ;
\label{3.12}
\\
\nonumber
\hat{\rho}_3^{(\rm out)}&=&\frac{2\alpha\beta}{N}
\left(\hat{\rho}^{({\rm in})}\right)^{\rm T}
+ \frac{(N-2\alpha\beta)}{N^2}
\hat{\openone} \, ,
\ee
where $\hat{\rho}^{({\rm in)}}=|\Psi\rangle \langle\Psi|$
is the density operator of the original state of system $1$,
and $\left(\hat{\rho}\right)^{\rm T}$ is the transposed
operator.

Taking into account condition (\ref{3.10}) we can directly rewrite
the last two density operators in the form (\ref{1.2}). 
This means that QID is the covariant transformation which
in a controlled way distributes information between the two systems.
There is a price to pay for this covariant information distribution
which is reflected by the additional noise.

\subsection{Cloner}
Let us assume that $\alpha=\beta$, i.e., the two outputs (\ref{3.12})
are equal. In this case  QID acts as a universal quantum cloner for
arbitrary dimensions. From (\ref{3.9}) we find the initial state of the 
cloner to be
\begin{eqnarray}
\label{3.13}
|\Phi\rangle_{23}=
\frac{1}{\sqrt{2(N+1)}}
  \sum_{m=0}^{N-1}\left(|x_0\rangle_2+|x_m\rangle_2\right)|x_m\rangle_3\;.
\end{eqnarray}
With this initial QID state the output of the cloner yields two clones of 
the form
\begin{eqnarray}
\label{3.14}
\hat{\rho}_{j}^{(\rm out)} = s \hat{\rho}_{j}^{({\rm in})} +
\frac{1-s}{N}\hat{\openone}\;,\qquad j=1,2\, .
\end{eqnarray}
The scaling factor $s$ is
\begin{eqnarray}
\label{3.15}
s=\frac{N+2}{2(N+1)}\;.
\end{eqnarray}
Finally, system $3$ of the cloner has a reduced state given by
\begin{eqnarray}
\label{3.16}
\hat{\rho}_{3}^{(\rm out)} = \frac{1}{N+1} 
\left(\hat{\rho}^{({\rm in})}\right)^{\rm T}
+ \frac{1}{N+1}\hat{\openone}\;, 
\end{eqnarray}
i.e., this piece of the cloner is left in a state proportional
to the transposed state of the original quantum system plus completely
random noise.

\section{Continuous limit}
\label{sec4}

In what follows we  make a connection between the discrete 
and the continuous case. The role of the controlled shifts (NOTs) 
in the continuous limit is obvious --- it is a conditional shift down 
the $x$-axis in phase space. Consequently, the QID operator (\ref{3.3}) 
has a clear meaning in the continuous limit.  Our goal now is to find the 
continuous analogue of the initial state
$|\Phi\rangle_{23}$ (\ref{3.9}) 
of the QID. This is rather straightforward: we simply need to use the 
regularized versions of the states $|x_0(\xi)\r$, $|p_0(\xi)\r$ and
$|\Xi_{00}(\xi)\r$ as introduced in Section~\ref{sec2}. 
The input state of the QID in the continuous case can then be written as 
\be
|\Phi(\xi)\r_{23}=
\alpha|\Xi_{00}(\xi)\r_{23}+\beta |x_0(\xi)\r_2 |p_0(\xi)\r_3\;.
 \label{4.1}
\ee
which in the $x$-basis becomes
\be
|\Phi(\xi)\r_{23}=
\frac{1}{2\pi}\int dx_2\, dx_3 \, \mu(x_2,x_3) |x_2\r|x_3\r\;,
 \label{4.2}
\ee
where 
\be
 \mu(x_2,x_3) = 
\alpha \psi_{|\Xi_{00}\r}(x_2,x_3)
+\beta \psi_{|x_{0}\r}(x_2)\psi_{|p_{0}\r}(x_3)\;,
 \label{4.3}
\ee
and the Gaussian functions 
$\psi_{|x_{0}\r}(x_2)$, $\psi_{|p_{0}\r}(x_3)$, 
and $\psi_{|\Xi_{00}\r}(x_2,x_3)$ are defined by 
Eqs.~(\ref{2.30}), (\ref{2.31}) and (\ref{2.35}), respectively.

For finite values of squeezing the states 
$|\Xi_{00}(\xi)\r_{23}$ and $|x_0(\xi)\r_2 |p_0(\xi)\r_3$ are not mutually
orthogonal, therefore, in order to fulfill the normalization
condition for the state $|\Phi\r$ the parameters
$\alpha$ and $\beta$ have to fulfill a condition analogous to (\ref{3.10})
\be
\alpha^2 + \beta^2 + \frac{4 \alpha \beta}{\sqrt{4 + 2\sinh^2 2\xi}}=1\;.
\label{4.4}
\ee

With this initial preparation of the QID, the universal (covariant)
information distribution of continuous variables is realized using the
network described in (\ref{3.2}). The operator $\hat{U}_{123}$ acts on 
the basis states $|x_1\r_1|x_2\r_2|x_3\r_3$ as
\be
\hat{U}_{123}|x_1\r_1|x_2\r_2|x_3\r_3=
|z_1\r_1|z_2\r_2|z_3\r_3\;,
\label{4.5}
\ee
where $z_1=x_1-x_2+x_3$, $z_2=x_1+x_2$, and $z_3=x_1+x_3$.
Assuming that the original system is initially prepared in the state
\be
|\Psi\r_1=\frac{1}{\sqrt{2\pi}}\int dx_1\, \psi(x_1) |x_1\r_1\;,
\label{4.6}
\ee
the output of the QID becomes
\end{multicols}
\vspace{-0.2cm}
\noindent\rule{0.5\textwidth}{0.4pt}\rule{0.4pt}{0.6\baselineskip}
\vspace{0.2cm}
\be
\hat{U}_{123}|\Psi\r_1|\Phi(\xi)\r_{23} =
\frac{1}{(2\pi)^{3/2}}\int dx_1\, dx_2 \, dx_3 \,
\psi(x_1) \mu(x_2,x_3)\,
|x_1-x_2+x_3\r_1\,|x_1+x_2\r_2\,|x_1+x_3\r_3\;.
\label{4.7}
\ee
Upon tracing out modes $2$ and $3$ we obtain from (\ref{4.7})
the density operator describing the original system at the output 
of the QID:
\be
\hat{\rho}_1^{(\rm out)}=
\frac{1}{(2\pi)^{3/2}}\int d\eta\, dx_1 \, dx_1' \,
\psi(x_1)\psi^*(x_1') {\cal K}(x_1-x_1';\eta)\,|x_1+\eta\r\l x_1'+\eta|\;,
\label{4.8}
\ee
where the integral kernel ${\cal K}(x_1-x_1';\eta)$ is given
by the expression
\be
{\cal K}(x_1-x_1';\eta)=\frac{1}{2\sqrt{2\pi}}\int d\chi \, 
\mu\Bigl(\frac{\chi-\eta}{2}-x_1;\frac{\chi+\eta}{2}-x_1\Bigr)\,
\mu\Bigl(\frac{\chi-\eta}{2}-x_1';\frac{\chi+\eta}{2}-x_1'\Bigr)\;.
\label{4.9}
\ee
From the fact that the trace of the density matrix (\ref{4.8}) is
equal to unity we find that the integral kernel (\ref{4.9})
has to fulfill the condition
\be
\frac{1}{\sqrt{2\pi}}\int d\eta\,
{\cal K}(0;\eta)=1\;.
\label{4.10}
\ee
The kernel itself can be expressed in the form
\be
{\cal K}(\bar{x}_1 ;\eta)=
\alpha^2 {\cal K}_1(\bar{x}_1 ;\eta)+
\beta^2 {\cal K}_2(\bar{x}_1 ;\eta) +
\alpha \beta {\cal K}_3(\bar{x}_1 ;\eta)\;,
\label{4.11}
\ee
where we have introduced the notation $\bar{x}_1=x_1-x_1'$.
Using the explicit expressions for the wave functions describing
the input state of the QID we find for the kernel functions
\be
{\cal K}_1(\bar{x}_1;\eta)&=&
e^{\xi}\exp\Bigl(-\frac{e^{-2\xi}}{2}\bar{x}_1^2
-\frac{e^{2\xi}}{2}\eta^2\Bigr)
\label{4.12} \\
{\cal K}_2(\bar{x}_1;\eta)&=&
\frac{1}{\sqrt{\cosh2\xi}}
\exp\Bigl(-\frac{\cosh2\xi}{2}\bar{x}_1^2
-\frac{1}{2\cosh2\xi}\, \eta^2\Bigr)
\label{4.13} \\
{\cal K}_3(\bar{x}_1;\eta)&=&
\frac{2}{\sqrt{3 e^{-2\xi}+e^{2\xi}}}
\exp\left[-\frac{
e^{-4\xi}(1+e^{4\xi})\bar{x}_1^2 + (2+ \sinh^22\xi)\eta^2}
{3 e^{-2\xi}+e^{2\xi}}\right]
\\
\nonumber
&&\times \left\{
\exp\left[-\frac{
e^{-4\xi}(1-e^{4\xi})\bar{x}_1}
{3 e^{-2\xi}+e^{2\xi}}\right] +
\exp\left[-\frac{
e^{-4\xi}(-1+e^{4\xi})\bar{x}_1}
{3 e^{-2\xi}+e^{2\xi}}\right]
\right\}\;.
\label{4.14}
\ee
  \hfill\noindent\rule[-0.6\baselineskip]%
  {0.4pt}{0.6\baselineskip}\rule{0.5\textwidth}{0.4pt}
\vspace{-0.2cm}
\begin{multicols}{2}
\noindent
It is now easy to check that
\be
\frac{1}{\sqrt{2\pi}}
\int d\eta\, {\cal K}_1(0;\eta)
&=&\frac{1}{\sqrt{2\pi}}
\int d\eta\, {\cal K}_2(0;\eta) =1
\nonumber
\\
\frac{1}{\sqrt{2\pi}}
\int d\eta\, {\cal K}_3(0;\eta)
&=&
\frac{4}{\sqrt{4+2\sinh^22\xi}} \;,
\label{4.15}
\ee
from which it follows that the kernel ${\cal K}(0;\eta)$
satisfies condition (\ref{4.10}).

In what follows we utilize the Wigner-function formalism to 
analyze the performance of the cloning machine. We find 
a Wigner $W(x,p)_1^{(out)}$ 
of the output state (\ref{4.8}) which we express as a convolution
of the Wigner function $W_1^{({\rm in})}(x,p)$ of the input mode
and the Wigner function $W^{{\cal K}}(x,p)$ of the kernel (\ref{4.11}):
\be
W(x,p)_1^{(out)} &=& \frac{1}{2\pi} \int dx' \, dp' 
W^{{\cal K}}(x',p'),
\nonumber
\\
&\times &
W^{({\rm in})}_1(x-x',p+p') 
\label{4.16}
\ee
where
\be
W^{{\cal K}}(x',p') = \frac{1}{\sqrt{2\pi}} \int
dz\, e^{ip' z} {\cal K}(z,x').
\label{4.17}
\ee
From our definitions it follows that 
\be
W^{{\cal K}}(x',p') &=& \alpha^2 W^{{\cal K}_1}(x',p')
+\beta^2 W^{{\cal K}_2}(x',p')
\nonumber
\\
&+&\alpha\beta W^{{\cal K}_3}(x',p')
\label{4.18}
\ee
where $W^{{\cal K}_j}(x',p')$ are the Wigner functions of the
kernels ${\cal K}_j$ ($j=1,2,3$).
We can easily check that 
\be
\frac{1}{2\pi}\int dx'\, dp'\, W^{{\cal K}}(x',p') =  1
\label{4.19}
\ee
which is equivalent to the condition (\ref{4.10}).

From Eq.(\ref{4.12}) we find the Wigner function $W^{{\cal K}_1}(x',p')$
\be
W^{{\cal K}_1}(x',p') = e^{2\xi} \exp\left[-\frac{e^{2\xi}}{2}(x'^2 + p'^2)
\right] \, ,
\label{4.20}
\ee
which in the limit of large squeezing reads
\be
W^{{\cal K}_1}(x',p') \rightarrow 2\pi \, \delta(x') \delta(p').
\label{4.21}
\ee
The Wigner function $W^{{\cal K}_2}(x',p')$ of the kernel (\ref{4.13})
reads
\be
W^{{\cal K}_2}(x',p') = \frac{1}{1+2\bar{n}} 
\exp\left[-\frac{(x'^2 + p'^2)}{2(1+2\bar{n})}
\right] \, ,
\label{4.22}
\ee
where we have used the notation $\bar{n}=\sinh^2 \xi$, so that
$\cosh 2\xi = 1+ 2 \bar{n}$.
We note that
this Wigner function for large squeezing (i.e. the large $\bar{n}$ limit)
is  equal to the Wigner function of a {\em thermal} state (\ref{2.34})
with the mean number of excitations equal to $2 \bar{n}$ !

Analogously we can evaluate the explicit expression for the Wigner function 
$W^{{\cal K}_3}(x',p')$ of the kernel (\ref{4.14}). This is rather 
cumbersome,
and, since we are interested only in the large squeezing limit,
we present the corresponding Wigner function only in this limit
\be
W^{{\cal K}_3}(x',p') \simeq 2 \sqrt{2}
\exp\left[-\frac{e^{2\xi}}{4}(x'^2 + p'^2)
\right] \, ,
\label{4.23}
\ee
which in the large $\xi$ limit can be formally expressed as
\be
W^{{\cal K}_3}(x',p') \rightarrow 8 \sqrt{2} \pi e^{-2\xi}
\delta(x') \, \delta(p').
\label{4.24}
\ee
Now we can give the explicit expression for the Wigner function, 
$W_1^{({\rm out})}(x,p)$, of the output mode 1 for which we find
\end{multicols}
\vspace{-0.2cm}
\noindent\rule{0.5\textwidth}{0.4pt}\rule{0.4pt}{0.6\baselineskip}
\vspace{0.2cm}
\be
W_1^{({\rm out})}(x,p) = \alpha^2 W_1^{({\rm in})}(x,p)
+\frac{\beta^2}{2\pi}
\int dx' \, dp' \, W_1^{({\rm in})}(x-x',p-p')
W^{{\cal K}_2}(x',p')
+ 4 \sqrt{2} e^{-2\xi} \alpha \beta W_1^{({\rm in})}(x,p)
\label{4.25}
\ee
  \hfill\noindent\rule[-0.6\baselineskip]%
  {0.4pt}{0.6\baselineskip}\rule{0.5\textwidth}{0.4pt}
\vspace{-0.2cm}
\begin{multicols}{2}

We note that in the large $\xi$ limit the third term in the
right-hand side of Eq.(\ref{4.25}) will vanish due to the factor
$e^{-2\xi}$. Taking into account that in the large squeezing 
(i.e. large $\bar{n}$) limit
the function $W^{{\cal K}_2}(x,p)$ is essentially equal to a 
Wigner function of a thermal field (\ref{2.34}) with a mean excitation
number of  $2 \bar{n}$ (we will denote this Wigner function
as $W_{\rm th}(x,p;2\bar{n})$), we can rewrite the Wigner function 
(\ref{4.25}) 
at the output of the QID as 
\be
\label{4.26}
&&W_1^{({\rm out})}(x,p) = \alpha^2 W_1^{({\rm in})}(x,p)
\\
&&+\frac{\beta^2}{2\pi}
\int dx' \, dp' \, W_1^{({\rm in})}(x',p')
W_{\rm th}(x-x',p-p';2\bar{n}).
\nonumber
\ee
Therefore, the output Wigner function is simply $\alpha^2$ times
the input Wigner function plus $\beta^2$ times the convolution
of the input Wigner function and that of a thermal state.

Finally we evaluate the fidelity of the QID transformation which 
is defined as
\be
{\cal F}_j&=&\l \Psi|\hat{\rho}_j^{(\rm out)}|\Psi\r 
\nonumber
\\
&=& 
\frac{1}{2\pi}\int dx\, dp\, W^{({\rm in})}_1(x,p) W^{({\rm out})}_j(x,p).
\label{4.32}
\ee
In the large squeezing limit the fidelity (\ref{4.32}) can be approximated
as
\end{multicols}
\vspace{-0.2cm}
\noindent\rule{0.5\textwidth}{0.4pt}\rule{0.4pt}{0.6\baselineskip}
\vspace{0.2cm}
\be
\int dx\, dp\, \int dx'\, dp'\ W^{({\rm in})}(x,p) 
W_{\rm th}(x-x',p-p',2\bar{n}) W^{({\rm in})}(x',p') \nonumber \\
\simeq \frac{2\pi}{2\overline{n}+1}\int dx\, \int dx'\, |\psi 
(x)|^{2}|\psi (x')|^{2}e^{-(x-x')^{2}/2(2\overline{n}+1)} .
\label{4.33}
\ee
  \hfill\noindent\rule[-0.6\baselineskip]%
  {0.4pt}{0.6\baselineskip}\rule{0.5\textwidth}{0.4pt}
\vspace{-0.2cm}
\begin{multicols}{2}
The integrals on the right-hand side of this equation are less
than or equal to $1$, so that the entire right-hand side goes
to zero as $1/\overline{n}$ as $\overline{n}\rightarrow \infty$.
Therefore, we find that in the large squeezing limit, the 
fidelity of the QID is indeed input-state
independent, and ${\cal F}_1=\alpha^2$ while 
${\cal F}_2=\beta^2$.

\subsection{Universal continuous cloner}

If in the limit $\xi\rightarrow\infty$ we take 
$\alpha^2 = \beta^2 =\case{1}{2}$ then the transformations described
above describe the symmetric cloner. 
In the limit $\xi\rightarrow\infty$, the
fidelity of the output density matrices 
to the original state is $\case{1}{2}$.
  This is consistent with what we expect from 
the limit $N\rightarrow \infty$ limit of Eq.~(\ref{3.14}).

The fact that in the continuous case the fidelity of the copies
is $\case{1}{2}$ is suggestive; it makes one think of a coin toss.  
In fact, one can construct a continuous (universal) cloner, which is much
simpler than the one given above but achieves the same fidelity,
whose most important component is a flipping coin 
\cite{Cerf2000}.  This cloner
has two inputs, one for the state we wish to clone and one for
a completely random state (ideally an infinite temperature thermal 
state). What the cloner does is to flip a coin, and if the result is 
heads, the original input state is sent to output $1$ and the random 
state to output $2$.  If instead the result is tails, the input state 
is sent to output $2$ and the random state to output $1$.  Assuming that, 
on average, the overlap between the input state and the random state 
is small, this ``cloner''
will clone the input with a fidelity of $\case{1}{2}$.  From this we
can conclude that for continuous quantum systems the universal
cloner is effectively a completely classical device. Indeed, one can
verify that in this limit there is no entanglement between systems $1$ and
$2$ of the outgoing particles. This is not true in any finite dimensional 
case.  Taking this classical distribution as a hint, we can see that 
this type of continuous cloner is easily generalized to the case of an 
arbitrary number of inputs, $M_{\rm in}$, and an arbitrary number of 
outputs, $M_{\rm out}$, with $M_{\rm out}\ge M_{\rm in}$.  In this case 
the fidelity of cloning is just $M_{\rm in}/M_{\rm out}$, which agrees 
with Werner's result \cite{werner} for the optimal cloner in the
infinite dimensional limit.      

\section{Conclusions}
\label{sec5}

We have shown that for any dimension, the optimal
universal quantum cloner can be constructed from essentially
the same quantum circuit, i.e., what we have is a universal
design for universal cloners.  In the case of infinite dimensions
(which includes continuous variable quantum systems) the universal cloner 
reduces to a classical device. By contrast, Cerf, et\ al., have shown 
that if one designs a continuous cloner optimized to copy certain 
sets of states, then one can achieve higher fidelities than those available
to the truly universal cloners studied here \cite{Cerf1999}.  In 
particular, they showed 
that it is possible to design a cloner that will copy any coherent 
state with a fidelity of $2/3$.  Their cloner also fits within the
structure of the QID analyzed here if one chooses the initial cloner state as
\begin{equation}
|\Phi\rangle_{23}=\frac{1}{\sqrt{2\pi^2}}\int\! dx_2 \,dx_3
\exp\Big(-\frac{x_2^2+x_3^2}{2}\Bigr)|x_2\r|x_2+x_3\rangle \;.
\label{5.1}
\end{equation}
It is interesting to note that this cloner also produces approximate
versions of the transpose of coherent states at its third
output.  The transpose of the coherent state $|z\r$ is 
$|z^{\ast}\r$.  If the input to the cloner is $|z\r$, then
$\hat{\rho}^{({\rm out})}_{3}$ is a Gaussian state (that is,
$\l x|\hat{\rho}^{({\rm out})}_{3}|x^{\prime}\r$ is a Gaussian) that
is concentrated about the point $z^{\ast}$ in phase space.  It
is, however, more spread out than a coherent state, and its
fidelity with the actual transposed state is $1/8$.

For continuous systems, these
specialized cloners will be more useful than the universal one.
Because of their covariance properties, they will clone any
two states that differ by only a translation in phase space with
the same fidelity.  For example, the fact that the cloner in
Ref.~\cite{Cerf1999} clones the vacuum with fidelity $2/3$ implies
that it clones all coherent states with the same fidelity.  One       
can easily imagine generalizing this result and designing cloners to
optimally clone entire classes of states; if by choosing the
correct $|\Phi\rangle_{23}$, the cloner has been optimized to
clone a particular state, $|\Psi\rangle$, by covariance it will 
automatically be optimal for all states generated from $|\Psi\rangle$ by
displacements in phase space.  For finite dimensional systems, however, 
universal cloners do better than classical devices and the simple 
universal circuit presented here shows how they may be constructed.  

\acknowledgements
We thank Gerard Milburn for helpful discussion.
This work was supported by the IST project EQUIP under the contract
IST-1999-11053 and by the National Science Foundation under grant 
PHY-9970507. V.B. acknowledges a support from the University of Queensland 
Traveling Scholarship. SLB was supported in part under project QUICOV 
under the IST-FET-QIPC programme.

\end{multicols}

\end{document}